\documentclass[11pt,fleqn,twoside]{article}
\usepackage{amsfonts,latexsym}
\makeatletter
\newcommand{\prava}{\footnotesize\it
\begin{flushright}
\begin{minipage}{18cm}
Copyright \copyright 1998 by Boris. A. Kupershmidt
\end{minipage}
\end{flushright}}

\newcommand{\name}[1]{\begin{flushleft}
                       \LARGE \bf #1
                       \end{flushleft}\vspace{-3mm}}

\newcommand{\Author}[1]{\begin{flushleft}
                       \it #1 \end{flushleft}}

\newcommand{\Adress}[1]{\begin{flushleft}
                       \it #1 \end{flushleft}}

\newcommand{\Date}[1]{\begin{flushleft}
                      \small  \it #1 \end{flushleft}}

\newcommand{\ehkol}{Author \ name}
\newcommand{\ohkol}{Article \ name}
\renewcommand{\@evenhead}{
\hspace*{-3pt}\raisebox{-15pt}[\headheight][0pt]{\vbox{\hbox to \textwidth
{\thepage \hfil \ehkol}\vskip4pt \hrule}}}
\renewcommand{\@oddhead}{
\hspace*{-3pt}\raisebox{-15pt}[\headheight][0pt]{\vbox{\hbox to \textwidth
{\ohkol \hfil \thepage}\vskip4pt\hrule}}}
\renewcommand{\@evenfoot}{}
\renewcommand{\@oddfoot}{}

\newcommand{\be}{\begin{equation}}
\newcommand{\ee}{\end{equation}}
\newcommand{\ba}{\hspace*{-5pt}\begin{array}}
\newcommand{\ea}{\end{array}}

\newcommand{\ds}{\displaystyle}
\makeatother

\font\BoldMath=cmmib10 scaled \magstep1
\font\BoldMathN=cmbsy10 scaled \magstep1

\newcommand{\pmal}{\mbox{\BoldMath \char 11}}
\newcommand{\pmba}{\mbox{\BoldMath \char 12}}
\newcommand{\pmlam}{\mbox{\BoldMath \char 21}}
\newcommand{\pmue}{\mbox{\BoldMath \char 22}}
\newcommand{\pmE}{\mbox{\BoldMath \char 69}}
\newcommand{\pmP}{\mbox{\BoldMath \char 80}}
\newcommand{\pmQ}{\mbox{\BoldMath \char 81}}
\newcommand{\pmX}{\mbox{\BoldMath \char 88}}
\newcommand{\pmd}{\mbox{\BoldMath \char 100}}
\newcommand{\pmp}{\mbox{\BoldMath \char 112}}
\newcommand{\pmq}{\mbox{\BoldMath \char 113}}
\newcommand{\pmu}{\mbox{\BoldMath \char 117}}
\newcommand{\pmna}{\mbox{\BoldMathN \char 114}}

\begin{document}

\setcounter{page}{162}

\thispagestyle{empty}

\renewcommand{\ehkol}{B.A. Kupershmidt}
\renewcommand{\ohkol}{Hamiltonian Formalism in Quantum Mechanics}

\begin{flushleft}
\footnotesize \sf
Journal of Nonlinear Mathematical Physics \qquad 1998, V.5, N~2,
\pageref{kupershmidt-fp}--\pageref{kupershmidt-lp}. \hfill {\sc Article}
\end{flushleft}

\vspace{-5mm}

\renewcommand{\footnoterule}{}
{\renewcommand{\thefootnote}{}
 \footnote{\prava}}

\name{Hamiltonian Formalism in Quantum Mechanics}\label{kupershmidt-fp}

\Author{Boris A. KUPERSHMIDT}

\Adress{The University of Tennessee Space Institute, Tullahoma, TN
37388  USA\\
E-mail: bkupersh@utsi.edu}

\Date{Received January 29, 1998}

\begin{flushright}
\begin{minipage}{6.5cm}
\small \bfseries \itshape To Vladimir Igorevich Arnol'd with
admiration, on occasion of his 60$^{\;th}$ birthday.
\end{minipage}
\end{flushright}

\begin{abstract}
\noindent
Heisenberg motion equations in Quantum mechanics can be put into the Hamilton form.  The dif\/ference between the commutator and its principal part, the Poisson bracket, can be accounted  for exactly.  Canonical transformations in Quantum mechanics are not, or at least not what they appear to be; their properties are formulated in a series of Conjectures.
\end{abstract}

\renewcommand{\thesection}{\arabic{section}}

\section{Introduction}
\setcounter{equation}{0}
\renewcommand{\theequation}{\arabic{section}.\arabic{equation}{\rm a}}

The motion equations of Classical mechanics, in the Hamilton form, are:\be
\dot q_i = {\partial H \over \partial p_i},
\ee

\setcounter{equation}{0}
\renewcommand{\theequation}{\arabic{section}.\arabic{equation}{\rm b}}
\be
\dot p_i = - {\partial H \over \partial q_i} .
\ee Here $i = 1,\ldots, N$, and $H$, the Hamiltonian, is a function of
the $p_i$'s and $q_i$'s, most often polynomial in the momenta
$p_i$'s.  The overdot, as usual, denotes the time derivative.
\setcounter{equation}{1}
\renewcommand{\theequation}{\arabic{section}.\arabic{equation}{\rm a}}

The motion equations of Quantum mechanics, in the Heisenberg form, are:\be
\dot q_i = h^{-1} [H, q_i],
\ee
\setcounter{equation}{1}
\renewcommand{\theequation}{\arabic{section}.\arabic{equation}{\rm b}}
\be
\dot p_i = h^{-1} [H, p_i].
\ee

\setcounter{equation}{2}
\renewcommand{\theequation}{\arabic{section}.\arabic{equation}}
Here $H$ again is a ``function'' of the $p_i$'s and $q_i$'s; the latter, however, {\it no  longer  commute} between themselves but are, instead, subject to the commutation relations\be
[p_k, q_\ell] = h \delta_{k \ell}, \qquad [p_k, p_\ell] = [q_k, q_\ell] = 0,
\ee the complex number $\sqrt{-1}$ having been absorbed into $h$ for
future
convenience.  The straight bracket notation stands for the commutator: \be
[u, v] = uv - vu.
\ee
These two types of motion equations are known as not entirely unrelated.  For example, if the $p_i$'s and the $q_i$'s are treated as operators, then the Classical equations (1.1) describe the motion of the mean values of these operators provided the Hamiltonian is quadratic in its arguments.  (This is a Corollary of Ehrenfest's Theorem.  These and other
mysteries are revealed in Messiah's classic text on Quantum Mechanics
[8].)

The f\/irst main result of this paper is an observation that the
Quantum motion equations (1.2) can be recast into the Classical form
(1.1) {\it provided one properly defines the notion of partial
derivatives} entering into the RHS of the equations (1.1).  This is
done in the next Section. The main idea is to treat Quantum notions as special instances of
noncommutative  objects and then utilize noncommutative algebra concepts.

If the motion equations (1.1) and (1.2) are rewritten in the equivalent form as, respectively,\be
\dot F = \{H, F\},
\ee \be
\dot F = h^{-1} [H, F],
\ee where $F$ is an arbitrary function of the $p_i$'s and $q_i$'s and
$\{\cdot, \cdot\}$  denotes the Poisson bracket:\be
\{H, F\} = \sum_i \left( {\partial H \over \partial p_i} {\partial F \over \partial q_i} - {\partial H \over \partial q_i} {\partial F \over \partial p_i} \right),
\ee one can ask whether these two forms are related in some precise manner.  Certainly, one knows that the Poisson bracket is the ``main part'' of the commutator, in the sense that\be
\lim_{h \rightarrow 0}  h^{-1} [H, F] = \{H, F\}, \ee
as a physicist would say, or\be
\{H, F\} = h^{-1} [H, F] \quad  (\mbox{mod} \ h)
\ee as is preferred by mathematicans.  In Section 3 we shall verify that, when the number $N$ of degrees of freedom equals 1, \be
h^{-1} [H, F] = {\sum_{s \geq 1}} {(-h)^{s-1} \over s!} \left( {\partial^s H \over \partial p^s} {\partial^s F \over \partial q^s} - {\partial^s F \over \partial p^s} {\partial ^s H \over \partial q^s} \right),
\ee where the partial derivatives in the RHS are understood in the same sense, to be def\/ined in Section 2, as those entering formulae (1.1) when considered noncommutatively.  (The general case $N \geq 1$ is covered by formula
(3.22).) We shall see that formula (1.10) is related to the
def\/inition of multiplication on the space of normally quantised Hamiltonians.

In Section 4 we consider the question of canonical transformations in Quantum mechanics, reformulate the Classical Jacobian conjecture into a symplectic object, quantize it, and state various generalizations of it.

\section{Heisenberg as Hamilton in disguise}
\setcounter{equation}{0}

Let us f\/irst f\/ix notations and conventions.  Our basic number f\/ield
${\cal{F}}$ (such as {\bf{Q}}, {\bf{R}}, {\bf{C}}, etc.) will be of characteristic zero; this is not essential for results, but allows shortcuts in proofs.  Instead of a f\/ield ${\cal F}$ one can take any associative ring (or {\bf{Q}}-algebra) commuting with the function-ring generators, but we shan't travel this route either, to avoid interruptions by remarks.  Our function rings will always be {\it polynomial}, again to bypass necessary pedantic comments; nothing much will change if we allow unspecif\/ied
 {\it functions} of the $q_i$'s (rational, algebraic, etc.) as is the case in practical mechanics, because all our formulae will describe identities between dif\/ferential operators, and the said identities remain true no matter what objects these dif\/ferential operators are allowed to act upon.

We start with the associative ring\be
C = C_u = {\cal F} \langle u_1,\ldots , u_m\rangle ,
\ee consisting of polynomials in {\it noncommuting variables}
$u_1,\ldots, u_m$; all with coef\/f\/icients in~${\cal F}$.
 (The coef\/f\/icients are always assumed to commute with the f\/ield variables $u_i$'s.) If $x \in C$ then $\hat
L_x$ and $\hat R_x$ denote the operators of left and right
multiplication by $x$ in $C$:\be
\hat L_x (y) = xy, \qquad  \ \hat R_x (y) = yx, \qquad \forall \ x, y \in C. \ee
The associative ring generated by the operators $\hat L_x$ and $\hat R_x$, for all $x$ in $C$, is denoted\be
Op_0 (C).
\ee We shall utilize the following useful elements in this operator ring ([7]):  For any $H \in C$, \be
{\partial^\sim H \over \partial u_k} \in Op_0 (C)
\ee
is the following operator:\be
{\partial^\sim H \over \partial u_k} (x) = {d \over d \epsilon} \bigg|_{\epsilon =0} \left(H\bigg|_{u_{k} \rightarrow u_{k} + \epsilon x} \right), \quad x \in C.
\ee Alternatively, we can describe the operation $\ds {\partial^\sim  \over \partial u_k}$ {\it itself} as a {\it derivation} (over ${\cal
F}$) of $C$ into $Op_0 (C)$:

\setcounter{equation}{5}
\renewcommand{\theequation}{\arabic{section}.\arabic{equation}{\rm a}}\be
{\partial^\sim \over \partial u_k} : \ C \rightarrow Op_0 (C), \ee
which acts on the generators of $C$ by the rule

\setcounter{equation}{5}
\renewcommand{\theequation}{\arabic{section}.\arabic{equation}{\rm b}}\be
{\partial^\sim \over \partial u_k} (u_s) = \delta_{ks}.
\ee
\setcounter{equation}{6}
\renewcommand{\theequation}{\arabic{section}.\arabic{equation}}

If $X \in \mbox{Der}(C)$ is a derivation of $C$ (over ${\cal F}$)
then, obviously,
\be
X(H) = \sum_k {\partial^\sim H \over \partial u_k} (X(u_k)) = : {\partial^\sim H \over \partial {\pmu}} ({\pmX}), \qquad
\forall \ H \in C. \ee

The same equality can be described in a more familiar form.  First, let us write suggestively, but imprecisely, \be
X =  \sum_k X (u_k) {\partial \over \partial u_k} , \qquad \forall \ X \in \mbox{Der} (C),
\ee
to mean nothing more than $X \in \mbox{Der}(C)$ is uniquely
determined by the action of $X$ on the $u_k$'s.  Second, let
\be
\Omega^1 (C) = \left\{ \sum_{ks} \varphi_{ks} du_k \psi_{ks} \ | \
\varphi_{ks}, \psi_{ks} \in C \right\}
\ee
be the $C$-bimodule of 1-forms over $C$, with the universal derivation $d: \ C \rightarrow \Omega^1 (C)$ acting naturally on the generators of $C$ by the rule\be
d (u_k) = du_k, \qquad k = 1,\ldots, m.
\ee
Then\be
d (H) = \sum_k {\partial^\sim H \over \partial u_k} (du_k),
\ee where $\ds {\partial^\sim H \over \partial u_k}$, as an element of
$Op_0 (C)$, is extended naturally to act on any $C$-bimodule, in this
case $\Omega^1 (C)$.  If we now def\/ine the familiar pairing\be
\Omega^1 (C) \times \mbox{Der} (C) \rightarrow C
\ee by the rule\be
\left\langle \sum \varphi_{ks} d u_k \psi_{ks}, X
\right\rangle = \left(\sum \varphi_{ks} du_k \psi_{ks}\right) (X) = \sum \varphi_{ks} X (u_k) \psi_{ks},
\ee then formula (2.7) can be rewritten in the familiar form\be
X (H) = \langle dH, X \rangle = dH (X).
\ee
So far we haven't met any $p$'s or $q$'s.  We shall get to them at the very end of this Section, for more general formulae we work with now are more transparent and easier to handle.

For lack of better notation, we shall denote by $\ds {\partial H
\over \partial u_k}$ the following element of the ring $C$, {\it not}
of the ring $Op_0 (C)$:\be
{\partial H \over \partial u_k} = {\partial^\sim H \over \partial u_k} (1) = {d \over d \epsilon} \bigg|_{\epsilon = 0} \left(H \bigg|_{u_k \mapsto u_k + \epsilon} \right).
\ee If $H$ is a homogeneous polynomial of degree $\ell = \mbox{deg}(H)$
and $X^{rad} \in \mbox{Der}(C)$ is the radial derivation of $C$:\be
X^{rad} (u_k) = u_k, \qquad k = 1,\ldots, m,
\ee
then we have the following noncommutative analog of the Euler Theorem on homogeneous functions:\be
X^{rad} (H) = \left(\sum u_k {\partial \over \partial u_k} \right) (H) = \ell H = \mbox{deg} (H) H.
\ee

Suppose now that we impose some {\it commutation  relations} on the $u_i$'s.  This means that we are given a f\/inite or inf\/inite system of polynomials (or, in more general circumstances, power series, etc.)
\setcounter{equation}{17}
\renewcommand{\theequation}{\arabic{section}.\arabic{equation}{\rm a}}
\be
R_r = \sum_\sigma c_{r \sigma} u^\sigma, \ \ \ c_{r \sigma}  \in {\cal F},
\ee
\setcounter{equation}{17}
\renewcommand{\theequation}{\arabic{section}.\arabic{equation}{\rm b}}
\be
u^\sigma: = u_{\sigma{_1}} \ldots u_{\sigma{_s}} \ \ \mbox{for} \ \ \sigma = (\sigma_1, \ldots,
\sigma_s) , \ \ \sigma_c = 1, \ldots, m,
\ee
and we form the factor-ring\setcounter{equation}{18}
\renewcommand{\theequation}{\arabic{section}.\arabic{equation}}
\be
C_u^{new} = C_u/I_{\cal R},
\ee where $I_{\cal R}$ is the two-sided ideal in $C_u$ generated by the
polynomials $R_r$'s.  If we want now to consider some ``motion
equations'' in the ring $C_u^{new}$, i.e., elements of the Lie algebra $\mbox{Der} (C_u^{new})$, we
have to look at only those derivations $X \in \mbox{Der} (C_u)$ which {\it
preserve the ideal} $I_{\cal R}$.  There exists quite a number of
such special derivations, namely the elements\be
\{\mbox{ad}_F : = \hat L_F - \hat R_F \ | \ F \in C_u\}.
\ee Indeed, any element of the ideal $I_{\cal R}$ is a f\/inite sum of the terms\be
\{ \varphi P_r \psi \ | \ \varphi, \psi \in C\}.
\ee But then\be
\mbox{ad}_F (\varphi P_r \psi) = F \varphi P_r \psi - \varphi P_r \psi F \ee
is again an element of $I_R$.  In the physical language, if\be
\dot u_i = [F, u_i], \qquad i = 1, \ldots, m,
\ee so that\be
u_i (t + \Delta t) = u_i (t) + \Delta t [F, u_i (t)] + O (\Delta t)^2,
\ee then\be
\sum c_{r \sigma} u^\sigma (t + \Delta t) = \sum c_{r \sigma} u^\sigma (t) + \Delta t \left[F, \sum c_{r \sigma} u^\sigma (t) \right]
+ O (\Delta t)^2, \ee
so that the commutation relations on the $u_i$'s are preserved in time.

There may exist also some other derivations of the ring $C_u$ which
preserve a particular ideal $I_{\cal R}$.  This is the case we are interested in, with the derivations in question being the ``partial derivatives''$\ds {\partial \over \partial u_k}$  (2.15).
\setcounter{equation}{26}

\medskip

{\samepage
\noindent
{\bf Lemma 2.26.}  {\it Suppose we are given the relations\be
P_{ij} = u_i u_j - u_j u_i - c_{ij}, \qquad c_{ij} = - c_{ij} \in
{\cal F}. \ee 
Then the derivations $\ds {\partial \over \partial u_k}$ preserve the
two-sided ideal generated by these relations.}}

\medskip

\noindent
{\bf Proof.}  We have\be
{\partial \over \partial u_k} (P_{ij}) = \delta_{ik} u_j + \delta_{jk} u_i - \delta_{jk} u_i - \delta_{ik} u_j = 0,
\ee and hence\be
{\partial \over \partial u_k} (\varphi P_{ij} \psi) = {\partial \varphi \over \partial u_k} P_{ij} \psi + \varphi P_{ij} {\partial \psi \over \partial u_k} \  \in I_{\cal R}. \qquad  \qquad \qquad \mbox{\rule{2mm}{4mm}}
\ee

\setcounter{equation}{30}

\noindent
{\bf Corollary 2.30.}  {\it In the ring $C_u^{new}$: \be
C_u^{new} = {\cal F} \langle  u_1,\ldots, u_m\rangle
 / \left([u_i, u_j] = c_{ij}\right)
\ee 
the objects\be
\left\{ {\partial H \over \partial u_k} \ \ \bigg| \ \ H \in
C^{new}_u, \quad k = 1, \ldots , m \right\}
\ee 
are well-defined and satisfy formulae\be
\mbox{\rm ad}_{u_{i}} (H) = \sum_k {\partial H \over \partial u_k } c_{ik}. \ee }

\noindent
{\bf Proof.}  By formula (2.7), \be
\mbox{ad}_{u_{i}} (H) = \sum_k {\partial ^\sim H \over \partial u_k}
(\mbox{ad}_{u_{i}}  (u_k)) \quad \mbox{in} \ C_u.
\ee By formula (2.27),\be
[u_i, u_j] = c_{ij} \quad  \mbox{in} \  C^{new}_u.
\ee 
Hence, now in $C^{new}_u$, \be
\mbox{ad}_{u_{i}} (H) = \sum_{k} {\partial^\sim H \over \partial u_k} (c_{ik}) = \sum c_{ik} {\partial^\sim H \over \partial u_k} (1) = \sum c_{ik} {\partial H \over \partial u_k}. \qquad \mbox{\rule{2mm}{4mm}}\ee 

\setcounter{equation}{37}

\noindent
{\bf Corollary 2.37.} {\it Consider the case where ${\cal F}$ is
replaced by
\be
{\cal F}_h = {\cal F} [\;[h]\;],
\ee 
the ring of formal power series in $h$, and the $u_i$'s are taken to be the $p_i$'s and the $q_i$'s, with the commutation  relations\be
[p_i, q_i] = h \delta_{ij}, \qquad
 [p_i, p_j] = [q_i, q_j] = 0.
\ee Then the Heisenberg motion equations (1.2) take the Hamiltonian form (1.1).}

\medskip

\noindent
{\bf Proof.}  We can transform formulae (1.2) as follows:\[
\dot q_i = h^{-1} [H, q_i] = - h^{-1} \mbox{ad}_{q_{i}} (H)
\ {\mathop {=}\limits^{[{\rm by} \ (2.33, 39)]}} \ - h^{-1} {\partial H \over \partial p_i} (-h) = {\partial H \over \partial p_i},
\]
\[
\dot p_i = h^{-1} [H, p_i] = - h^{-1} \mbox{ad}_{p_{i}} (H) = - h^{-1} {\partial H \over \partial q_i} h = - {\partial H \over \partial q_i}.\qquad \qquad \qquad \mbox{\rule{2mm}{4mm}}
\]
\setcounter{equation}{40}

\noindent
{\bf Remark 2.40.} Like in the commutative algebra and analysis, partial derivatives commute between themselves both in $C_u$ and $C_u^{new}$:\be
{\partial ^2 H \over \partial u_i \partial u_j} \ = \ {\partial ^2 H \over \partial u_j \partial u_i}.
\ee This is clear from the def\/inition (2.15).
\setcounter{equation}{42}

\medskip

\noindent
{\bf Remark 2.42.}  The {\it operator-valued} partial derivatives $\ds
{\partial^\sim H \over \partial u_k}$satisfy the chain rule:
If the $u_k$'s are functions of the $\varphi_\alpha$'s then \be
{\partial ^\sim H \over \partial \varphi _\alpha} = \sum_k {\partial ^\sim H \over \partial u_k} {\partial ^\sim u_k \over \partial \varphi_\alpha} .
\ee Indeed, \be
u_k (\varphi_1,\ldots, \varphi_\alpha + \epsilon x,\ldots) =
u_k (\varphi) + \epsilon {\partial^\sim u_k \over \partial \varphi_\alpha} (x) + O
\left(\epsilon ^2\right).
\ee Therefore, \be
\ba{l}
\ds H (u_1 (\varphi _\alpha + \epsilon x), \ldots, u_m (\varphi_\alpha + \epsilon x)) = H \left(u_1 (\varphi) + \epsilon {\partial^\sim u_1 \over \partial \varphi_\alpha} (x) + O \left(\epsilon^2\right), \ldots
\right) \\[4mm] \ds \qquad =
H (u (\varphi)) + \epsilon \sum_k {\partial ^\sim H \over \partial u_k} \left({\partial^\sim u_k \over \partial \varphi_\alpha} (x) \right), \ea
\ee 
so that\be
{d \over d \epsilon} \bigg|_{\epsilon =0} H (\varphi_1,\ldots, \varphi_\alpha+ \epsilon x,\ldots) = {\partial^\sim H \over \partial \varphi_\alpha} (x) = \sum {\partial^\sim H \over \partial u_k} {\partial^\sim u_k \over \partial \varphi_\alpha} (x) , \qquad  \forall \ x,
\ee 
and formula (2.43) follows.

\section{Commutator vs Poisson bracket}
\setcounter{equation}{0}

On the way to verify formula (1.10), we shall prove f\/irst a more general statement.
Suppose we impose the relations\be
[u_i, u_j] = h c_{ij}, \qquad  1 \leq i,  \ j \leq m,
 \quad  c_{ij} = - c_{ij}  \in  {\cal F},
\ee on the ring ${\cal F}_h \langle u_1, \ldots , u_m\rangle$.  We can
think of these relations as the rules allowing us to reduce every
polynomial in the $u_i$'s to a specif\/ic lexicographic form by choosing an ordering among the generators $u_i$'s.  The original relations (3.1), in the form\be
u_i u_j = u_j u_i + h c_{ij},
\ee imply, and are equivalent to, the series of relations\be
{u_i^n \over n!} {u^m_j \over m!} = \sum_{s \geq 0} (h c_{ij})^s {u_j^{m-s} \over (m-s)!} {u_i^{n-s} \over (n-s)!} , \qquad n , m \in
{\bf N}. \ee 
This series of relations, in turn, is equivalent to the {\it single} formal relation
\be
E^{\pmlam \cdot \pmu} E^{\pmue \cdot \pmu} = E^{\pmue \cdot \pmu} E^{\pmlam\cdot \pmu} {e^{h \langle {\pmlam}, {\pmue} \rangle }}
\ee 
in ${\cal F}_h \langle u_1,\ldots, u_m\rangle [[{\pmlam}$, ${\pmue}]]$, where\be
E^{\pmlam \cdot \pmu} = e^{\lambda_1 u_1}\ldots
e^{\lambda_m u_m},
\ee \be
\langle {\pmlam}, \pmue \rangle = \sum c_{ij} \lambda_i \mu_j =
- \langle \pmue, {\pmlam}\rangle.\ee 

\setcounter{equation}{7}

\noindent
{\bf Lemma 3.7.}  {\it Define the coefficients $\{\theta_{\sigma \sigma^{\prime}}\}$ in ${\cal F}$ by the identity:\be
\sum_{s \geq 1} {(-h)^{s-1} \over s!}
\langle  \pmlam, \pmue \rangle^s = \sum_{\sigma \sigma^\prime} \theta_{\sigma \sigma^\prime} \lambda^\sigma \mu^{\sigma\prime}{(-h)^{-1+(|\sigma|+|\sigma^\prime|)/2} \over ((|\sigma| + |\sigma^\prime| )/2)! },
\ee where
\setcounter{equation}{8}
\renewcommand{\theequation}{\arabic{section}.\arabic{equation}{\rm a}} 
\be
\lambda^\sigma = \lambda^{\sigma_{1}}_1 \ldots \lambda^{\sigma_{m}}_m \quad\mbox{for} \ \ \sigma = (\sigma_1,\ldots, \sigma_m), \quad \sigma_i
\in {\bf Z}_+,
\ee \setcounter{equation}{8}
\renewcommand{\theequation}{\arabic{section}.\arabic{equation}{\rm b}}
\be
|\sigma| = \sigma_1 + \cdots  + \sigma_m.
\ee 
\setcounter{equation}{9}
\renewcommand{\theequation}{\arabic{section}.\arabic{equation}}

\noindent
Then\be
h^{-1} [H, F] = \sum_{s \geq 1}{(-h)^{s-1} \over s!} \sum_{|\sigma | = |\sigma^\prime|=s} \theta_{\sigma \sigma^{\prime}} {\partial^{|\sigma|} H \over \partial u^\sigma} {\partial^{|\sigma^\prime|} F \over \partial u^{\sigma^{\prime}}},
\ee where\be
{\partial^{|\sigma|}H \over \partial u^\sigma} =
{\partial^{\sigma_{1}} \over \partial u^{\sigma_{1}}_1} \cdots
{\partial^{\sigma_{m}} \over \partial u_m^{\sigma_{m}}} (H) \qquad  \mbox{for} \quad
 \sigma = (\sigma_1,\ldots, \sigma_m).
\ee }

\noindent
{\bf Proof.}  It's enough to check formula (3.10) for the case\be
H = E^{\pmlam \cdot \pmu}, \qquad
 F = {E^{{\pmue} \cdot \pmu}}.
\ee 
Then\[
\ba{l}
\ds h^{-1} [H, F] = h^{-1}
\left(\pmE^{\pmlam \cdot \pmu} \pmE^{\pmue \cdot \pmu} - \pmE^{\pmue \cdot \pmu} \pmE^{\pmlam \cdot \pmu} \right)
\\[3mm]
\ds \qquad \ {\mathop {=} \limits^{\rm [by \ (3.4)]}} \ \ HF {1 - e^{h \langle  \pmue, \pmlam \rangle } \over h} =
HF {1 - e^{-h \langle  \pmlam,  \pmue \rangle } \over h}
 = {e^{-h \langle  \pmlam, \pmue\rangle } - 1 \over -h} HF \\[3mm]\ds \qquad = \sum\limits_{s \geq 1} {(-h)^{s-1} \over s!}
\langle  \pmlam, \pmue\rangle ^s HF \ \ {\mathop {=}\limits^{\rm [by \ (3.8)]}} \ \
 \sum \theta_{\sigma \sigma^\prime} \lambda^\sigma \mu^{\sigma^{\prime}} {(-h)^{-1+\cdots} \over (\cdot \cdot \cdot) !} \\[3mm]\ds \qquad= \sum\limits_{s \geq 1} {(-h)^{s-1} \over s!}
\sum\limits_{|\sigma| = |\sigma{^\prime}|=s}
\theta_{\sigma \sigma^{\prime}} {\partial^{|\sigma|} H \over \partial u^\sigma} {\partial^{|\sigma^{\prime} |} F \over \partial u^{\sigma^{\prime}}}. \qquad  \qquad \qquad \qquad \qquad \mbox{\rule{2mm}{4mm}}
\ea
\]

The terms with $s=1$ in the RHS of formula (3.10) comprise the
Poisson bracket part, for formula (3.8) implies that\be
\theta_{ij} = c_{ij}.
\ee 
If we now specialize to the Quantum case when $[p_i, q_j] = h \delta_{ij}$, we {\it will  not} get formula (1.10), for in the RHS of formula (3.10) $H$ stands always to the {\it left} of $F$ and the $(H, F)$ -- skewsymmetry is thus hidden.  But we can emulate the {\it proof} of
Lemma 3.7.  First, we  convert the relations\be
[p_i, q_j] = h \delta_{ij}, \qquad  [p_i, p_j] = [q_i, q_j] = 0,
\ee into the singl formal relation\be
e^{\pmlam \cdot \pmp} e^{\pmal \cdot \pmq} = e^{h \pmlam \cdot \pmal} e^{\pmal \cdot \pmq} e^{\pmlam \cdot \pmp}.
\ee Next, we take\be
H = e^{\pmal \cdot \pmq} e^{\pmlam \cdot \pmp}, \qquad
F = e^{\pmba \cdot \pmq}  e^{\pmue \cdot \pmp}.
\ee Now, consider the operators\setcounter{equation}{16}
\renewcommand{\theequation}{\arabic{section}.\arabic{equation}{\rm a}}
\be
{\cal O}_{HF} = {\partial^H \over \partial \pmp} \cdot {\partial^F \over \partial \pmq} : \ HF \mapsto \sum_i {\partial H \over \partial p_i} {\partial F \over \partial q_i},
\ee \setcounter{equation}{16}
\renewcommand{\theequation}{\arabic{section}.\arabic{equation}{\rm b}}
\be
{\cal O}_{FH} = {\partial^F \over \partial \pmp} \cdot {\partial^H \over \partial \pmq} : \ FH \mapsto \sum_i {\partial F \over \partial p_i} {\partial H \over \partial q_i}.
\ee \setcounter{equation}{17}
\renewcommand{\theequation}{\arabic{section}.\arabic{equation}}
Let us verify that\be
e^{-h {\cal O}_{HF}} (HF)= (\mbox{smbl} (H) \mbox{smbl}
(F))_{\mbox{\footnotesize normal}},
\ee 
where, for $H\in C^{new}_{u}$, $\mbox{smbl}(H) \in {\cal F}
[u_1,\ldots, u_m]$ is the symbol of $H$ which results by letting $h$ vanish (in $C^{new}_{u} / (hC_u^{new})),$ and the subscript ``normal'' denotes the normal quantization, with the $q_i$'s standing to the left of the $p_i$'s.  Indeed, for $H$ and $F$ given by formula (3.16), \be
e^{-h {\cal O}_{HF}} (HF) = \sum_{s \geq 0} {(-h \pmlam \cdot \pmba)^s \over s!} HF = e^{(\pmal + \pmba) \cdot \pmq} e^{(\pmlam + \pmue) \cdot \pmp}. \ee 
Since\be
\mbox{smbl} (H) \mbox{smbl} (F)  =
\mbox{smbl} (F) \mbox{smbl} (H),
\ee formula (3.18) implies:\be
e^{-h {\cal O}_{HF}} (HF) = e^{-h {\cal O}_{FH}} (FH),
\ee 
so that\be
\ba{l}
\ds h^{-1} [H, F] = h^{-1} (HF - FH) = {e^{-h {\cal O}_{HF}} -1  \over -h} (HF) - {e^{-h {\cal O}_{FH}}-1 \over -h} (FH) \\[3mm]\ds \qquad \qquad = \sum\limits_{s \geq 1} {(-h)^{s-1} \over s!}
\left(({\cal O}_{HF})^s (HF) - ({\cal O}_{FH})^s (FH) \right).
\ea
\ee For the case when the number of degrees of freedom $N=1$, \be
{\cal O}_{HF} (HF) = {\partial H \over \partial p} {\partial F \over
\partial q} ,
\ee and formula (3.22) yields formula (1.10).
\medskip

\setcounter{equation}{24}

\noindent
{\bf Remark 3.24.} Formula (3.21) implies that we have a symmetric bilinear form in the {\it noncommutative} ring  $C^{new}_{p,q}:$ \be
(H, F) = e^{-h {\cal O}_{HF}} (HF).
\ee There exists another attractive bilinear form on this ring, this time with values in the {\it commutative} ring ${\cal F} [ q_i] [[h]]$: \be
(H, F) = \mbox{Res} (HF^\dagger),
\ee where\be
\mbox{Res} \left(\sum c_{\sigma \sigma^{\prime}} q^\sigma
p^{\sigma^{\prime}} \right) : = \sum c_{\sigma 0} q^\sigma,
\ee and $\dagger$ is an antiinvolution (over ${\cal F}_h)$:\be
(HF)^\dagger = F^\dagger H^\dagger,
\ee def\/ined on the generators $q_i$'s and $p_i$'s by the rule \be
q^\dagger_i = q_i, \qquad  p_i^\dagger = - p_i, \qquad
 i = 1, \ldots, N.
\ee 
The bilinear form $(H,F)$ (3.26) is not symmetric in the {\it linear algebra} sense, but it is symmetric in the {\it differential  algebra} sense:\be
(H, F) \sim (F, H),
\ee where, for elements $a, b \in {\cal F} [q_i] [\;[h]\;]$, we write \be
a \sim b \quad \mbox{to  mean} \quad (a - b) \in \sum_i Im {\partial \over \partial q_i}.
\ee To prove formula (3.30), we take $H$ and $F$ given by formula (3.16).  Then\be
\ba{l}
(H, F) = \mbox{Res} (HF^\dagger) = \mbox{Res} \left(
e^{\pmal \cdot \pmq} e^{\pmlam \cdot \pmp} e^{- \pmue \cdot \pmp} e^{\pmba \cdot \pmq} \right) \\[2mm]\qquad =\mbox{Res} \left(e^{(\pmal + \pmba) \cdot \pmq}
e^{(\pmlam - \pmue) \cdot \pmp} e^{(\pmlam - \pmue ) \cdot \pmba h} \right)=
e^{(\pmlam - \pmue) \cdot \pmba h}e^{(\pmal + \pmba) \cdot \pmq}.
\ea
\ee 
Hence,\be
\ba{l}
\ds (F, H) = e^{(\pmue - \pmlam) \cdot \pmal h} e^{(\pmal + \pmba) \cdot \pmq} = e^{(\pmue - \pmlam) \cdot (\pmal + \pmba ) h} (H, F) \\[2mm]\qquad = e^{h (\pmue - \pmlam) \cdot \partial/\partial \pmq} (H, F) \sim (H, F).
\ea
\ee 
\setcounter{equation}{0}

\section{Canonical transformations, special and general}

If $M$ is a smooth manifold and $T^*M$ is the contangent bundle (= the phase space) of $M$, then any transformation \be
\varphi : \ M \rightarrow M
\ee is uniquely lifted to a transformation\be
\bar \varphi : T^* M \rightarrow T^* M
\ee covering $\varphi$, by the requirement that the canonical 1-form\be
\rho = \pmp d \pmq
\ee on $T^*M$ be preserved:\be
\bar \varphi^* (\rho) = \rho.
\ee Re-expressing this picture analytically/algebraically, we start with an automorphism $\Phi$ of the ring $C_q$\be
\Phi: C_q \rightarrow C_q, \qquad C_q = {\cal F} [q_1,\ldots q_N],
\ C^\infty (q_1,\ldots, q_N), \ \ldots
\ee \be
\Phi (q_i) = Q_i = Q_i (q_1,\ldots,q_N), \qquad i = 1,\ldots, N,
\ee and then determine the elements\be
\bar \Phi (p_i) = P_i = P_i (q, p)
\ee from the requirement that \be
\pmp \pmd \pmq = \pmP \pmd \pmQ:
\ee \be
\sum_j p_j dq_j = \sum_i P_i d Q_i = \sum P_i Q_{i,j} dq_j. \ee 
Thus,\be
p_j = \sum_i P_i Q_{i,j}, \qquad j = 1,\ldots, N.
\ee Denote by\be
J = J_{Q|q} = \left({\partial Q_i \over \partial q_j} \right)
\ee the Jacobian of the map $\Phi$.  The transformation formulae
(4.10) can be rewritten in one of the equivalent forms:
\setcounter{equation}{11}
\renewcommand{\theequation}{\arabic{section}.\arabic{equation}{\rm a}}
\be
\pmp^t = \pmP^{\;t} J,
\ee 
\setcounter{equation}{11}
\renewcommand{\theequation}{\arabic{section}.
\arabic{equation}{\rm b}}\be  \pmP^{\;t} = \pmp^t J^{-1},
\ee \setcounter{equation}{11}
\renewcommand{\theequation}{\arabic{section}.\arabic{equation}{\rm c}}
\be
\pmP = (J^{-1})^t \pmp,
\ee where $\pmp$, $\pmP$, $\pmq$, $\pmQ$ are thought of as
column-vectors.  Since the canonical 1-form $\pmp d \pmq$ is
preserved, the symplectic 2-form $d \pmp  \wedge d \pmq$ is preserved as well.  Therefore, the basic Poisson brackets are also preserved:\setcounter{equation}{12}
\renewcommand{\theequation}{\arabic{section}.\arabic{equation}}
\be
\{ P_i, P_j\} = \{Q_i, Q_j\} = 0, \qquad \{P_i, Q_j\} = \delta_{ij}.
\ee 
\setcounter{equation}{14}

\noindent
{\bf Remark 4.14.}  If one concentrates on the preservation of the Poisson brackets {\it only}, that is, of the 2-form $d \pmq \wedge d \pmq$, rather than the canonical 1-form $\pmp d \pmq$, the uniqueness of the lifting of $\varphi$ into $\overline \varphi$ no longer holds.  For example, we can replace formula (4.4) by the relation\be
\overline \varphi^{\;*} (\rho) = \rho + \omega,
\ee where $\omega$ is a closed 1-form on $M$ lifted into $T^*M$.  Taking\be
\omega = d (f), \qquad f \in C_q,
\ee we f\/ind, instead of formula (4.10), the relations\be
p_j = \sum P_i Q_{i,j} + f,_j,
\ee \be
\pmP = (J^{-1})^t \left(\pmp - \pmna (f)\right).
\ee We shall see  below that such nonuniqueness is {\it{unavoidable}} in Quantum mechanics.
\medskip

\setcounter{equation}{19}

\noindent
{\bf Lemma 4.19.}  {\it Formulae (4.6,12c) preserve the Quantum commutation relations\be
[q_i, q_j] = [p_i, p_j] = 0, \qquad  [p_i, q_j] = h \delta_{ij},
\quad 1 \leq i, j \leq N.
\ee }

\noindent{\bf Proof.}  Obviously,\be
[Q_i, Q_j] = 0.
\ee Next,\be\hspace*{-10.1pt}
[P_i, Q_j] = \left[\sum_\alpha \left(J^{-1}\right)^t_{i \alpha}
p_\alpha, \ Q_j
\right]  = \sum \left(J^{-1}\right)_{\alpha i} h Q_{j, \alpha} = h \sum_\alpha
\left(J^{-1}\right)_{\alpha i}J_{j \alpha} = h \delta_{ij}.
\ee Finally,\be\hspace*{-10.1pt}
P_i P_j = \sum \left(J^{-1}\right)_{\alpha i} p_\alpha
\left(J^{-1}\right)_{\beta j} p_\beta = \sum \left(J^{-1}\right)_{\alpha i}
\left\{\left(J^{-1}\right)_{\beta j} p_\alpha + h \left(J^{-1}\right) _{\beta j, \alpha} \right\} p_\beta,
\ee whence\be
h^{-1} [P_i, P_j ] = \sum_\beta \left\langle \sum_\alpha \left(J^{-1}\right)_{\alpha i} \left(J^{-1}\right)_{\beta j, \alpha} - \sum_\nu \left(J^{-1}\right)_{\nu j} \left(J^{-1}\right)_{\beta i, \nu} \right\rangle p_\beta.
\ee Now,\setcounter{equation}{24}
\renewcommand{\theequation}{\arabic{section}.\arabic{equation}{\rm a}}
\be
\left(J^{-1}\right)_{\beta j, \alpha} = -
\left(J^{-1} J,_\alpha J^{-1}\right)_{\beta j} = - \sum_{\mu \nu} \left(J^{-1}\right)_{\beta \mu} J_{\mu \nu, \alpha}
\left(J^{-1}\right)_{\nu j},
\ee 
and thus
\setcounter{equation}{24}
\renewcommand{\theequation}{\arabic{section}.\arabic{equation}{\rm b}}\be\left(J^{-1}\right)_{\beta i, \nu} = - \sum_{\mu \alpha}
\left(J^{-1}\right)_{\beta \mu} J_{\mu \alpha, \nu} \left(J^{-1}\right)_{\alpha i}.
\ee Substituting formulae (4.25) into formula (4.24) and noticing that
\setcounter{equation}{25}
\renewcommand{\theequation}{\arabic{section}.\arabic{equation}}
\be
J_{\mu \nu, \alpha} = {\partial^2 Q_\mu \over \partial q_\nu \partial q_\alpha} = J_{\mu \alpha, \nu} ,\ee 
we f\/ind that\be
[P_i, P_j] = 0. \qquad \qquad \qquad \qquad \mbox{\rule{2mm}{4mm}}
\ee 
The nonuniqueness of quantum formulae (4.12c) can be demonstrated in two ways.
\medskip

\setcounter{equation}{28}

\noindent
{\bf Lemma 4.28.} {\it  The transformation\be
Q_i = q_i, \qquad P_i = p_i + g,_i, \qquad i = 1,\ldots, m, \quad
g \in C_q, \ee 
preserves the quantum commutation relations (4.20).  }
\medskip

\noindent
{\bf Proof.}  We have\be
[P_i, P_j] = [p_i + g,_i, \ p_j + g,_j] = hg,_{ij} - h g,_{ji} = 0,
\ee and the rest of the relations are obviously satisf\/ied. \qquad
\rule{2mm}{4mm}
\medskip

\setcounter{equation}{31}
\renewcommand{\theequation}{\arabic{section}.\arabic{equation}{\rm a}}

\noindent
{\bf Lemma 4.31.}  {\it The transformation\be
Q_i = Q_i (\pmq),
\ee 
\setcounter{equation}{31}
\renewcommand{\theequation}{\arabic{section}.\arabic{equation}{\rm b}}\be
P_i = \sum_\alpha p_\alpha \left(J^{-1}\right)_{\alpha_{i}} , \qquad
i = 1,\ldots, N,
\ee is also a quantum canonical transformation.}

\setcounter{equation}{32}
\renewcommand{\theequation}{\arabic{section}.\arabic{equation}}

\medskip

\noindent
{\bf Proof.}  (A) \ The new formulae (4.32) are just the mirror image of the old ones, (4.6,12c), and $\dagger$ is an
(anti)isomorphism.
\noindent
(B) \ Alternatively, we can straightforwardly calculate like in the proof of Lemma 4.19, and keep all the $p_\alpha$'s to the {\it left} of the $q_\beta$'s. \qquad \rule{2mm}{4mm}

Thus, given a transformation $\Phi: \ C_q \rightarrow C_q$, we have two {\it different} lifts of it into quantum canonical maps, $\Phi_r$
(4.6,12c),   and $\Phi_\ell$ \ (4.32):\be
\Phi_r (q_i) = Q_i (\pmq), \qquad \Phi_r (p_i) = \sum_\alpha \left(J^{-1}\right)_{\alpha i} p_\alpha,
\ee \be
\Phi_\ell  (q_i) = Q_i (\pmq), \qquad \Phi_\ell (p_i) = \sum_\alpha p_\alpha \left(J^{-1}\right)_{\alpha i}.
\ee How are these two maps related?  Let us consider the composition $\Psi = \Phi_r \Phi^{-1}_{\ell}:$\be
\Psi (q_i) = q_i, \qquad \Psi (p_i) = \sum_{\alpha \beta}
\left(J^{-1} \right)_{\alpha \beta} p_\alpha J_{\beta i}, \qquad 1 \leq i \leq N.
\ee 
\setcounter{equation}{36}
\renewcommand{\theequation}{\arabic{section}.\arabic{equation}{\rm a}}

\noindent
{\bf Lemma 4.36.} {\it\be
\Psi (\pmp) = \pmp + h \pmna (g),
\ee \setcounter{equation}{36}
\renewcommand{\theequation}{\arabic{section}.\arabic{equation}{\rm b}}
\be
g = \ln\det (J).
\ee }
\setcounter{equation}{37}
\renewcommand{\theequation}{\arabic{section}.\arabic{equation}}

\noindent
{\bf Proof.}  From formulae (4.35) we f\/ind:\be
\Psi (p_i) = \sum \left(J^{-1}\right)_{\alpha \beta} \left\{ J_{\beta
i} p_\alpha + h J_{\beta i, \alpha} \right\} = p_i + h \sum
\left(J^{-1}\right)_{\alpha \beta} J_{\beta i, \alpha} .
\ee But \be
\ba{l}
\ds \sum \left(J^{-1}\right)_{\alpha \beta}
J_{\beta i, \alpha} = \sum \left(J^{-1}\right)_{\alpha \beta} J_{\beta \alpha, i} = \mbox{Tr} \left(J^{-1} J,_i\right) \\[3mm]
\ds \qquad \qquad \qquad \qquad \qquad {\mathop {=}\limits^{\rm {[by
\ formula \ (4.42) \ below]}}} \  (\ln \det (J)),_i. \qquad \qquad \qquad \mbox{\rule{2mm}{4mm}}
\ea
\ee 
\setcounter{equation}{40}

\noindent
{\bf Remark 4.40.} Recall that if $A \in  \mbox{Mat}_n (C_q)$ then\be
d (\ln \det (A)) = \mbox{Tr} (A^{-1} d A),
\ee in the sense that\be
(\ln \det (A)),_i = \mbox{Tr} (A^{-1} A,_i) = \sum (A^{-1})_{\alpha
\beta} A_{\beta \alpha, i}.
\ee Indeed, Let $B = \ln (A)$, so that $A = e^B$.  Then\[
\ba{l}
\ds \mbox{Tr} (A^{-1} dA) = \mbox{Tr} \left(e^{-B} d(e^B)\right) =
\mbox{Tr} \left(e^{-B} \sum {B^s d(B) B^r \over (r + s + 1)!} \right) \\[4mm]
\ds \qquad =  \mbox{Tr} \left(e^{-B} \sum {B^r B^s d(B) \over (r+s+1)!} \right)
 = \mbox{Tr} \left(e^{-B} \sum {B^\ell d(B) \over \ell !} \right)
= \mbox{Tr} (dB)) \\[4mm]
\ds \qquad = d\; \mbox{Tr} (B) = d (\ln \det (e^B)) = d (\ln \det(A)).
\ea
\]
\setcounter{equation}{43}

\noindent
{\bf Remark 4.43.}  Formula (4.41) is rational in $A$.  An equivalent formulation, regular in $A$, is \be
d (\det (A)) = \mbox{Tr} (\mbox{adj} (A) dA),
\ee 
where $\mbox{adj}(A) $ is the adjugate matrix of $A$:  \be
\mbox{adj} (A) A = A\; \mbox{adj} (A) = \det (A) {\bf{1}}.
\ee 
\setcounter{equation}{46}

\noindent
{\bf Remark 4.46.}  Which one of the maps $\Phi_r$ or $\Phi_\ell$ is
{\it right} in practice?  Unfortunately, this is the sort of question
akin to the problem of ``right'' quantization, that is to say, a wrong
and misleading one.  The ``right'' answer depends on the problem at
hand, i.e., the Hamiltonian, and it may be nonunique nonetheless.  I
shall leave an elaboration of this point to
the future.  Let us consider instead an instructive case of the {\it mechanical} Hamiltonians, those of the form\be
H = \sum a^{ij} (Q) P_i P_j + V (Q).
\ee It is well-known in Quantum mechanics that if $P_k$'s are  treated as $\ds h {\partial \over \partial Q_k}$'s (recall that $\sqrt{-1}$ has been absorbed into $h$) then the selfadjoint form of $H$ is\be
H = \sum P_i a^{ij} (Q) P_j + V (Q).
\ee In other words, \be
H^\dagger = H.
\ee How does one transform such an $H$ under a change of variables $q_i \mapsto \Phi (q_i) = Q_i (\pmq)$ \ (4.6), and still preserve the
selfadjointness of $H$?  Let us look at the simple example of a free  particle in polar coordinates:
\setcounter{equation}{49}
\renewcommand{\theequation}{\arabic{section}.\arabic{equation}{\rm a}}
\be
x = r\cos \theta, \ y = r\sin \theta,
\ee \setcounter{equation}{49}
\renewcommand{\theequation}{\arabic{section}.\arabic{equation}{\rm b}}
\be
J = \left(\matrix{\cos \theta & - \sin \theta \cr\sin \theta & \cos \theta \cr} \right) \left(\matrix{ 1 & 0 \cr0 & r \cr} \right) \Rightarrow J^{-1} = \left(\matrix{1 & 0 \cr 0& r^{-1} \cr} \right) \left(\matrix{\cos \theta & \sin \theta \cr-\sin \theta &
\cos \theta \cr} \right).
\ee Thus, for the left form (4.34) we get\setcounter{equation}{50}
\renewcommand{\theequation}{\arabic{section}.\arabic{equation}}
\be
\ba{l}
(p_x, p_y) = (p_r, p_\varphi) \left(\matrix{1 & 0 \cr0 & r^{-1} \cr} \right) \left(\matrix{\cos \theta & \sin \theta \cr-\sin \theta & \cos \theta \cr} \right) \\[4mm]\qquad \qquad = (p_r \cos \theta - p_\theta r^{-1} \sin \theta, \ p_r \sin
\theta + p_\theta  r^{-1} \cos \theta).
\ea
\ee Hence, for the right form (4.33) we obtain\be
(p_x, p_y) = \left(\cos \theta p_r - r^{-1} \sin \theta p_\theta,
\sin \theta p_r + r^{-1} \cos \theta p_\theta\right).
\ee Now, the Hamiltonian $p_x^2 + p_y^2$ becomes:\setcounter{equation}{52}
\renewcommand{\theequation}{\arabic{section}.\arabic{equation}$\ell$}
\be
p^2_x + p^2_y = p^2_r + r^{-2} p^2_\theta - h p_r r^{-1} \qquad\mbox{(left  form)},
\ee \setcounter{equation}{52}
\renewcommand{\theequation}{\arabic{section}.\arabic{equation}$r$}
\be
p_x^2 + p_y^2 = p_r^2 + r^{-2} p^2_\theta + hr^{-1} p_r \qquad\mbox{(right  form)},
\ee and neither of these is physically palatable by virtue of not being
selfadjoint.  This observation seems to suggest that a substantial
fraction of literature on Quantum mechanics is beside the point.
What the point or points is or are I'll again leave for the
future can-of-worms operations.  Let us return to the mechanical
Hamiltonian $H$ (4.48): how {\it should} it transform in order to
preserve its selfadjointness?  We have seen above that neither the
left nor the right transformation is satisfactory.
\setcounter{equation}{54}
\renewcommand{\theequation}{\arabic{section}.\arabic{equation}}

\medskip

\noindent
{\bf Lemma 4.54}.  {\it Denote the left and right transformations as\be
P_i^\ell = \sum p_\alpha \left(J^{-1}\right)_{\alpha i},
\qquad P^r_i = \sum \left(J^{-1} \right)_{\alpha i} p_\alpha .
\ee Set\be
H^{\ell r} = \sum P^\ell_i a^{ij} (Q) P_j^r , \qquad H^{r \ell} = \sum P^r_i a^{ij} (Q) P^\ell_j.
\ee Then\be
(H^{\ell r})^\dagger = H^{\ell r} , \qquad (H^{r \ell})^\dagger =
H^{r \ell}. \ee 
}

\noindent
{\bf Proof.}\be
(P_i^\ell)^\dagger = - P_i^r, \qquad (P_i^r)^\dagger = - P_i^\ell.
\qquad \qquad \qquad \mbox{\rule{2mm}{4mm}}
\ee 
In coordinates,\be
H^{\ell r} = \sum p_\alpha \left(J^{-1}\right)_{\alpha i}
a^{ij} (Q) \left(J^{-1}\right)_{\beta j}  p_\beta,
\ee \be
H^{r \ell} = \sum \left(J^{-1}\right)_{\alpha i} p_\alpha a^{ij} (Q) p_\beta \left(J^{-1}\right)_{\beta j}.
\ee One can try other remedies, e.g.\be
P_i  = (P_i^\ell + P_j^r)/2, \qquad P_i = \sqrt{P_i^\ell P_i^r} ,
\qquad P_i = \sqrt{P_i^r P_i^\ell},\ee 
etc., but they appear too artif\/icial.  It does seem unavoidable to work with {\it two different} type of momenta in Quantum mechanics, left and right, and transform each one accordingly.  Formula (4.59) appears to of\/fer slight advantages in this regard.  In particular, for the free  particle in polar coordinates, we f\/ind
\setcounter{equation}{61}
\renewcommand{\theequation}{\arabic{section}.\arabic{equation}{\rm a}}\be
H^{\ell r} = p_x^\ell p_x^r + p_y^\ell p_y^r = p^2_r + r^{-2} p^2_\theta, \ee 
\setcounter{equation}{61}
\renewcommand{\theequation}{\arabic{section}.\arabic{equation}{\rm b}}
\be
H^{r \ell} = p_x^r p_x^\ell + p_y^r p_y^\ell = p_r^2 + r^{-2} p_\theta^2, \ee 
and each one of these formulae is satisfactory.  In general,\setcounter{equation}{62}
\renewcommand{\theequation}{\arabic{section}.\arabic{equation}}
\be
\ba{l}
\ds h^{-2} (H^{\ell r} - H^{r \ell}) = h^{-2} \sum \left(
\left( J^{-1}\right)_{\alpha i} p_\alpha + h \left(J^{-1}\right)_{\alpha i, \alpha}\right)
 a^{ij} \left(J^{-1}\right)_{\beta j} p_\beta \\[3mm]\ds \qquad -h^{-2} \sum \left(J^{-1}\right)_{\alpha i} p_\alpha a^{ij}
\left(\left(J^{-1}\right)_{\beta j} p_\beta + h \left(J^{-1}\right) _{\beta j, \beta}\right) \\[3mm]\ds \qquad = h^{-1} \sum \left(J^{-1}\right)_{\mu j, \mu} a^{ij}
\left(J^{-1}\right)_{\beta i} p_\beta - h^{-1} \sum \left(J^{-1}\right)_{\beta i} p_\beta  a^{ij}
\left(J^{-1}\right)_{\mu j, \mu} \\[3mm]\ds \qquad = h^{-1} \sum \left[ \left(J^{-1}\right)_{\mu j, \mu}
a^{ij}, \left(J^{-1}\right)_{\beta i} p_\beta \right] \\[3mm]
\ds \qquad = - h^{-1} \sum \left(J^{-1}\right)_{\beta i} h
\left(\left(J^{-1}\right)_{\mu j, \mu} a^{ij}\right), _\beta \\[3mm]
\ds \qquad {\mathop {=}\limits^{\rm [by \ 4.64)]}} \ \sum \left(J^{-1}\right)_{\beta i}  \left((\ln \det(J)),_{\psi}
\left(J^{-1}\right) _{\psi j} a^{ij} \right),_\beta ,
\ea
\ee where we used the formula\be
\ba{l}
\ds - \sum\limits_\mu \left(J^{-1}\right)_{\mu j, \mu} = \sum
\left(J^{-1} J,_\mu J^{-1}\right)_{\mu j} = \sum \left(J^{-1}\right)_{\mu \varphi} J_{\varphi \psi, \mu}
\left(J^{-1}\right)_{\psi j} \\[3mm]\ds \qquad \qquad \qquad {\mathop{=}\limits^{\rm [by \ (4.42)]}} \ \sum\limits_\psi (\ln \det (J)),_\psi \left(J^{-1}\right)_{\psi j}.
\ea
\ee 
Let us return now to the formula (4.37).  It can be looked at from a slightly dif\/ferent perspective, if we notice that $\Psi (\pmp) = \pmp$ whenever\be
\det (J) = \mbox{const} \not= 0.
\ee Namely, from formulae (4.33,34) we f\/ind that\be
\ba{l}
\ds \Phi_r (p_i) - \Phi_\ell (p_i) = \sum \left(J^{-1}\right)_{\alpha
i} p_\alpha - p_\alpha \left(J^{-1}\right)_{\alpha i}) =
\sum \left[\left(J^{-1}\right)_{\alpha i}, p_\alpha\right]
\\[3mm] \ds \qquad = - h \sum \left(J^{-1}\right)_{\alpha i, \alpha} \ \
{\mathop {=}\limits^{\rm [by \ (4.64)]}} \ \ = h \sum \left(J^{-1}\right)_{\alpha i} (\ln \det (J)),_\alpha .
\ea
\ee Thus, the condition of constant $\det (J)$ (4.65) is necessary and suf\/f\/icient to have the left and right formulae coincide, and thus provide a {\it unique} lift from an automorphism $\Phi$ of $C_q$ into a Quantum automorphism $\overline \Phi$ of $C_{p,q}$.   That a polynomial map $\Phi$ with a constant non-zero $\det (J)$ does indeed def\/ine an automorphism of $C_q$, has been conjectured originally by Keller in [5]; this conjecture is known as the Jacobian Conjecture, and it is related to many other
open problems in algebra; see , e.g., reviews in [2].  Let us discuss this Conjecture, thereafter called Conjecture $K$, from the physical point of view.
First, since an automorphism of ${\cal F}[q]$
extends, via formulae (4.12), to a Poisson automorphism of ${\cal F} [q, p]$,  Conjecture  $K$ is implied by the more general symplectic
\medskip

\noindent
{\bf Conjecture {\itshape S}.}  A polynomial Poisson endomorphism of ${\cal F} [p, q]$ is an automorphism.  (In other words, if $P_i, Q_i \in {\cal F} [p, q]$ are such that\be
\{P_i, P_j\} = \{Q_i, Q_j\} = 0, \qquad  \{P_i, Q_j\} = \delta_{ij}, \qquad  1 \leq i, j \leq N,
\ee then the $p_i$'s and the $q_i$'s can be re-expressed as
polynomials in
the $P$'s and the $Q$'s.  In other words still, ${\cal F} [P, Q] = {\cal
F}[p, q]$.)

Vice versa, the symplectic Conjecture $S$ is implied by the Conjecture
$K$.
Indeed, if the form $d \pmp \wedge d \pmq$ is preserved then so is
the volume form $(d \pmp \wedge \pmq)^{\wedge N}$; thus, the $\det (J)$ in this case equals to 1.

The symplectic Conjecture $S$ is a quasiclassical limit of the Quantum

\medskip
\noindent
{\bf Conjecture {\itshape Q}.}  Let $W_N = W_N (k;h)$ be the $h$-scaled  Weyl algebra over a commutative ring $k$, (see , e.g., [1]) with the
generators $q_1,\ldots, q_N$, $p_1,\ldots, p_N$ and the relations \be
[q_i, q_j] = [p_i, p_j] = 0, \qquad [p_i, q_j] = h \delta_{ij}, \qquad1 \leq i, j \leq N.
\ee 
If $Q_1,\ldots, Q_N$,  $P_1,\ldots, P_N \in W_N$ are such that\be
[Q_i, Q_j] = [P_i, P_j] = 0, \qquad [P_i, Q_j] = h \delta_{ij},
\qquad  1 \leq  i, j \leq N,
\ee then the $q_i$'s and $p_i$'s can be re-expressed as polynomials in the $P$'s and $Q$'s (with coef\/f\/icients in $k[h]$ or $k_h = k[[h]]$ depending upon the version of $W_N$).

In-between Conjectures $S$ and $Q$ is located
\medskip

\noindent
{\bf Conjecture {\itshape C--Q}.} (i) \ Every Poisson endomorphism (resp. automorphism) of ${\cal F}[p, q]$ can be quantized;
\noindent (ii) \ Such quantization  is unique over $k[h]$.

Quantization is certainly nonunique over $k[[h]]$.  For example,\be
P_i = \psi (h) p_i, \qquad  Q_i = q_i /\psi (h), \qquad i = 1,\ldots, N, \ee 
is a quantum automorphism for any $\psi (h) \in 1 + h k [[h]]$, and it reduces to an identical Poisson map for $h = 0$ no matter what $\psi (h)$ is; such nonuniqueness, therefore, attaches to {\it every} Quantum endomorphism.  On the other hand, results of Wollenberg [9, 3] show that the (i) part of the Conjecture $C$--$Q$ fails for {\it
infinitesimal} endomorphisms, i.e., derivations.  (See  also the last Remark at the end of this Section.)

Conjectures $K$ and $S$ have noncommutative analogs.
\medskip

\noindent
{\bf Conjecture  ${\cal K}$.}  Let $R$ be an associative ring and $R \langle x\rangle= R \langle x_1,\ldots, x_m \rangle$ a ring of
polynomials in noncommuting variables $x_1,\ldots, x_m$ with
coef\/f\/icients in $R$ which {\it do not} necessarily commute with the
$x_i$'s.  Let $F_1,\ldots, F_m \in R
\langle x\rangle$ be such that the Jacobian matrix $J(F)$:\be
J (F)_{ij} = {\partial ^\sim F_i \over \partial x_j} \  \in \ Op_0 (R \langle x\rangle)
\ee is invertible, so that there exists a matrix ${\cal M} \in
\mbox{Mat}_m (Op_0 (R\langle x\rangle ))$ such that\be
{\cal M} J (F) = {\bf{1}}.
\ee Then there exist polymonials $G_1,\ldots, G_m \in R\langle x\rangle $
such that \be
G_i (F_1,\ldots, F_m) = x_i, \qquad  1 \leq i \leq m.
\ee 

\noindent
{\bf Conjecture ${\cal S}$.} In $R \langle p,q\rangle  = R
\langle p_1,\ldots , p_N, q_1,\ldots, q_N\rangle$, let the
noncommutative Hamiltonian structure [7] be given by the Hamiltonian matrix\be
B = \left(\matrix{{\bf{0}} & {\bf{1}} \cr- {\bf{1}} & {\bf{0}} \cr} \right) .
\ee If $P_1,\ldots, P_N$, $Q_1,\ldots, Q_N \in R \langle p, q\rangle $
preserve the Hamiltonian structure $B$:\be
J B J ^\dagger = B,
\ee then $p_1,\ldots, p_n$, $q_1,\ldots, q_N \in R \langle P, Q
\rangle$. (The adjoint $J^\dagger $ of $J$ in formula (4.75) is taken in the noncommutative sense def\/ined in [7].)

We conclude this Section by mentioning two other versions of
Conjecture ${\cal K}$:

\medskip

\noindent
{\bf Conjecture ${\cal K}^{var}$.}  If $R$ is commutative and $F_1,\ldots, F_m \in R\langle x\rangle$ are such that the matrix
$J^{var} (F) \in\mbox{Mat}_m (R\langle x\rangle)$ [7] is invertible, where \be
J^{var} (F)_{ij} = {\delta F_i \over \delta x_j} \ \in \ R\langle
x\rangle,
\ee \be
{\delta F_i \over \delta x_j} \chi \equiv {\partial^\sim F_i \over \partial x_j} (\chi) \ (\mbox{mod} \; [R\langle x\rangle, R
\langle x\rangle ]), \qquad   \forall \ \chi \in R \langle x\rangle, \ee 
then the map $F:R^m \rightarrow R^m$ is a polynomial automorphism.
\medskip

\noindent
{\bf Conjecture $C$--${\cal K}$.} (A noncommutative analog of the Quantization Conjecture $C$--$Q$).  Let $f: k^m \rightarrow k^m$ be a polynomial map with an invertible Jacobian (resp. automorphism).  Then one can f\/ind a set of polynomials $F_1,\ldots, F_m \in k\langle x\rangle$
with an invertible Jacobian  (in either of the two meanings, (4.71) or (4.76)) (resp. automorphism) such that $f_i (x) = F_i (x)$, $i = 1,\ldots, m$, when
all the $x_i$'s are allowed to commute between themselves.

The Conjecture $C$--${\cal K}$ is obviously true for tame
automorphisms (generated by $GL_m (k)$ and triangular maps), and thus is true for $m=2$ ([4, 6]).  The same conclusion applies to the (i) part of Conjecture $C$--$Q$ for the case $N=1$.

\label{kupershmidt-lp}

The * attached to a citation marks the publisher that demands, as a
condition of printing a paper in the publisher's journal, that all
authors of creative works surrender and hand in to the publisher the
copyright to the fruits of their labors. The publishers so noted do
not include those in Cuba,
North Korea, and other savage places, where such a policy is not a
matter of free choice but is state-mandated.

\end{document}